\documentclass[]{spie}  

 \usepackage[T1]{fontenc}
\usepackage{lmodern}
\usepackage{amsmath,amsfonts,amssymb}
\usepackage{graphicx}
\usepackage{sidecap}
\usepackage{wrapfig}
\usepackage{subfig}
\usepackage[colorlinks=true, allcolors=blue]{hyperref}

\usepackage{afterpage}

\usepackage{siunitx}
\DeclareSIUnit{\cmps}{\cm\per\second}
\DeclareSIUnit{\mps}{\meter\per\second}
\DeclareSIUnit{\kmps}{\kilo\meter\per\second}
\DeclareSIUnit{\micrometer}{\micro\meter}
\DeclareSIUnit{\foot}{'}
\DeclareSIUnit{\inch}{"}

\title{On-sky commissioning of MAROON-X: A new precision radial velocity spectrograph for Gemini North}

\author[a]{Andreas Seifahrt}
\author[a]{Jacob L. Bean}
\author[a,b]{Julian St\"urmer}
\author[a]{David Kasper}
\author[c]{Luke Gers}
\author[d]{Christian Schwab}
\author[e]{Mathias Zechmeister}
\author[f]{Gu\dh mundur Stef\'ansson}
\author[g]{Ben Montet}
\author[a,h]{Leonardo A. Dos Santos}
\author[i]{Alison Peck}
\author[i]{John White}
\author[i]{Eduardo Tapia}
\affil[a]{The University of Chicago, Chicago, United States}
\affil[b]{Landessternwarte Heidelberg, Heidelberg, Germany}
\affil[c]{W. M. Keck Observatory, Kamuela, United States}
\affil[d]{Macquarie University, Sydney, Australia}
\affil[e]{Georg-August-Universit\"at G\"ottingen, G\"ottingen, Germany}
\affil[f]{Princeton University, Princeton, United States}
\affil[g]{University of New South Wales (UNSW), Sydney, Australia}
\affil[h]{Observatoire de l’Universit\'e de Gen\`eve, Versoix, Switzerland}
\affil[i]{Gemini Observatory, Hilo, United States}

\authorinfo{Further author information: E-mail: seifahrt@uchicago.edu}

\begin{document} 
\maketitle

\begin{abstract}
MAROON-X is a fiber-fed, red-optical, high precision radial velocity spectrograph recently commissioned at the Gemini North telescope on Mauna Kea, Hawai'i. With a resolving power of 85,000 and a wavelength coverage of 500--920\,nm, it delivers radial velocity measurements for late K and M dwarfs with sub-50\,cm\,s$^{-1}$ precision. MAROON-X is currently the only optical EPRV spectrograph on a 8\,m-class telescope in the northern hemisphere and the only EPRV instrument on a large telescope with full access by the entire US community. We report here on the results of the commissioning campaign in December 2019 and early science results. 
\end{abstract}


\section{INTRODUCTION}
\label{sec:intro}  
Our team at The University of Chicago has recently commissioned a new red-optical (500 -- 920\,nm), high-resolution (R\,$\simeq$\,85,000) radial velocity (RV) spectrograph (named ``MAROON-X'') that was designed explicitly for the purpose of following up small transiting planets around mid to late M dwarfs with sub-m\,s$^{-1}$ RV precision. 

MAROON-X was installed at the Gemini North Observatory in May 2019 and saw first light in September 2019. About 20\, hours of commissioning and science verification observations were performed in December 2019. The first year of operations was impacted by the COVID-19 pandemic which did not allow for site visits in order to complete all aspects of the commissioning. Notwithstanding, regular science observations have started in May 2020 and we conducted them remotely from Chicago. Despite the incomplete commissioning, a number of successful observing runs allow us to report on important performance figures of the instrument, including the capability to obtain sub-m\,s$^{-1}$ precision RVs in the first few months of the instrument's operation. 

Although MAROON-X is currently classified as a Visiting Instrument\footnote{See \url{https://www.gemini.edu/instrumentation/current-instruments/maroon-x} for more information.}, it is essentially permanently installed at Gemini North. The instrument was first offered to the Gemini community in the 2020B Call for Proposals, and demand for the instrument skyrocketed with the 2021A CfP.

In the following sections we provide a brief overview of the project timeline, update the as-built specifications of the instrument, discuss its current performance, including an example of an early scientific result highlighting the excellent RV precision delivered by MAROON-X, and present our plans for further improvements in 2021. 

\section{The MAROON-X Project}
The timeline of the project is shown in Table~\ref{tab:timeline}. The schedule was mainly driven by the availability of funds.

\begin{table}[h]
\caption{MAROON-X project timeline}
\begin{center}
\begin{tabular}{|l|l|}
\hline
2014/06	& Project PDR / kick-off meeting \\
2015/06	& Spectrometer contract signed with KiwiStar Optics\\
2015/08	& FDR for Spectrometer\\
2016/07 & FP etalon shows $<$3\,cm\,s$^{-1}$ precision\\
2016/08 & Pupil slicer first light in the lab\\
2017/01	& Spectrometer acceptance Chicago (one arm only)\\
2017/06 & First solar spectra taken with MAROON-X\\
2017/06	& PDR for telescope frontend (aka fiber injection unit)\\
2017/11	& FDR for telescope frontend\\
2018/07	& Environmental chamber installed in Pier Lab at Gemini North\\
2018/10	& Science-grade CCD detector systems received in Chicago\\
2018/10	& Red arm acceptance and integration in Chicago\\
2018/12	& Frontend commissioned on Gemini North\\
2019/06	& Spectrograph delivered and installed at Gemini North\\
2019/09	& First light on sky\\
2019/12	& Commissioning and science verification\\
2020/05 & First science observations\\
\hline
\end{tabular}
\end{center}
\label{tab:timeline}
\vspace{0mm}
\end{table}

\begin{table}[!b]
\caption{MAROON-X main characteristics}
\begin{center}
\begin{tabular}{|r|l|}
\hline
Spectral resolution & R $\simeq$ 85,000 \\
Acceptance angle & FOV = 0.77'' at the 8\,m Gemini North Telescope \\
Wavelength range & 500 nm -- 920 nm (in 56 orders)\\
Number and reach of arms & 2 (500--670\,nm and 650--920\,nm) \\
Cross-disperser & anamorphic VPH grisms\\
Beam diameter & 100\,mm (at echelle grating), 33\,mm (at cross-disperser)\\
Main fiber & 100\,$\mu$m octagonal (CeramOptec)\\
Number and type of slicer & 3x pupil slicer \\ 
Slit forming fibers & 5$\times$ 50$\times$150\,$\mu$m rectangular (CeramOptec)\\
Inter-order and inter-slice spacing & $\geq10$ pixel \\
Average sampling & 3.5 pixel per FWHM\\
Blue detector & Standard epi 30\,$\mu$m thick 4k$\times$4k STA4850 CCD (15\,$\mu$m pixel size)\\
Red detector & Deep-depletion 100\,$\mu$m thick 4k$\times$4k STA4850 CCD (15\,$\mu$m pixel size)\\
Calibration & Fabry-P\'erot etalon for simultaneous reference (fed by 2nd fiber) \\
Environment for main optics & Vacuum operation, 1\,mK P-V temperature stability goal\\
Environment for camera optics & Pressure sealed operation, 20\,mK P-V temperature stability\\
Short-term instrument stability & \SI{<0.3}{\mps} (demonstrated)\\
Long-term instrument stability & \SI{0.7}{\mps} (requirement), \SI{0.5}{\mps} (goal)\\
Total efficiency & 8\% at peak (demonstrated) at \SI{700}{\nano\meter} (at median seeing)\\
Observational efficiency & S/N$\simeq$90 per pixel at \SI{820}{\nano\meter} for a V=16.5 late M dwarf in \SI{30}{\minute} \\
\hline
\end{tabular}
\end{center}
\label{tab:parameters}
\vspace{0mm}
\end{table}

\subsection{Instrument overview}
\begin{figure}[t!]
\centering
\includegraphics[width=0.99\linewidth]{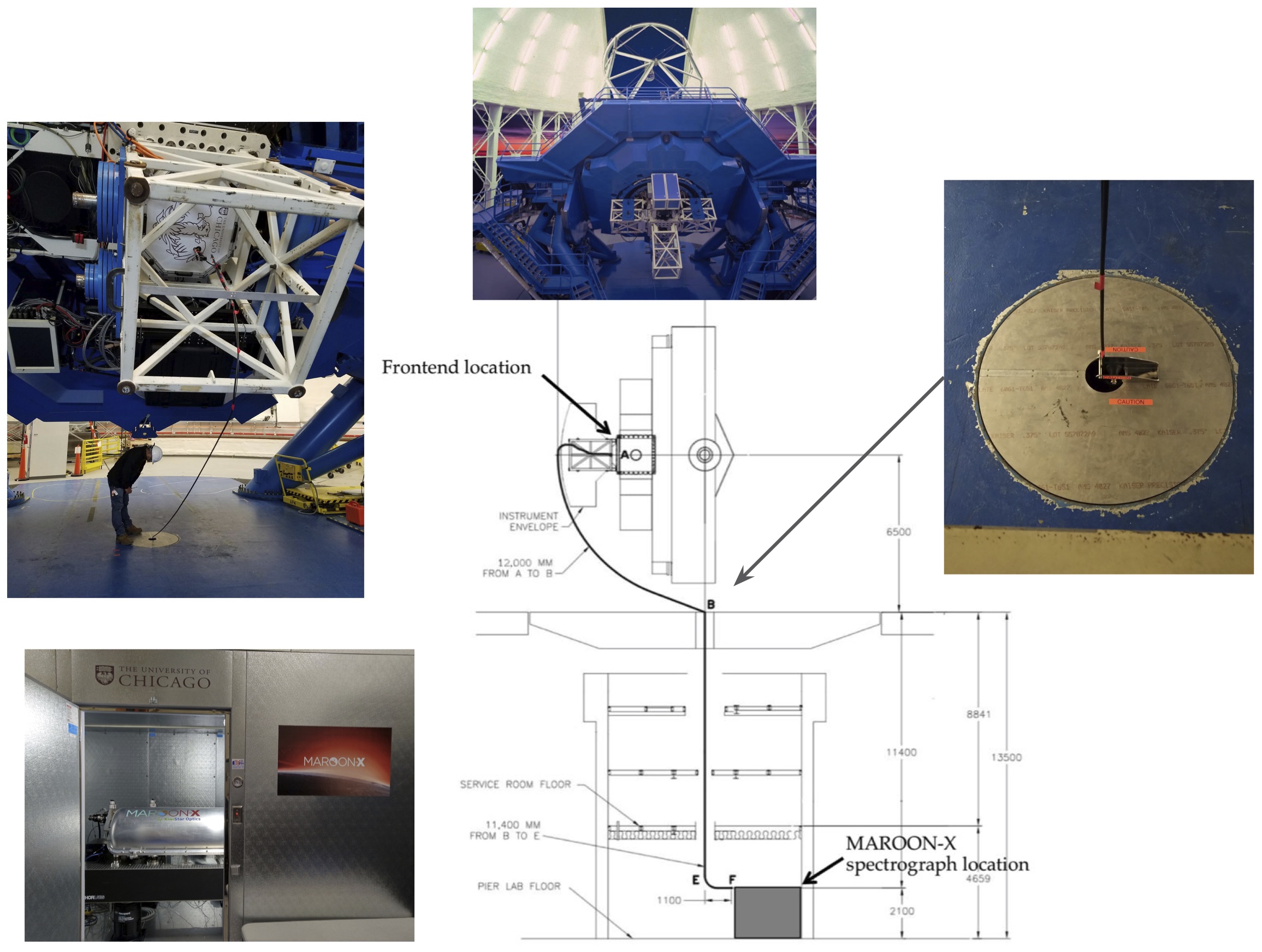}
\vspace{2mm}
\caption{\textbf{Location of MAROON-X components at the Gemini North Telescope.} The telescope frontend, mounted at the Instrument Support Structure (ISS) at the bottom Cassegrain port of the telescope, is feeding stellar light through an ADC and $f$/\#\,-\,conversion optics into the science fiber. The science fiber is contained in a fiber conduit that runs from the backend of the ISS through a central hole underneath the telescope straight down to the Pier Lab and into the spectrograph enclosure.}
\label{fig:overview}
\end{figure}

The MAROON-X spectrograph is installed inside an actively controlled walk-in enclosure located in the Pier Lab at ground level underneath the Gemini North Telescope, the same location the GHOST spectrograph\cite{GHOST} will occupy at Gemini South. Stellar light is injected into a \SI{100}{\micrometer} octagonal fiber in the telescope frontend (aka Fiber Injection Unit - FIU) and routed to the spectrograph in a fiber conduit that runs from the Instrument Support Structure (ISS) through a hole underneath the telescope straight down to the Pier Lab (see Fig.~\ref{fig:overview}). This setup allows for a realtively short fiber run of $\sim$30\,m between the telescope frontend and the spectrograph. An updated list of the main characteristics of MAROON-X is shown in Table~\ref{tab:parameters}.

\subsection{Fiber link}
\label{sec:fiber}
\begin{figure}[t!]
\centering
\includegraphics[width=0.49\linewidth]{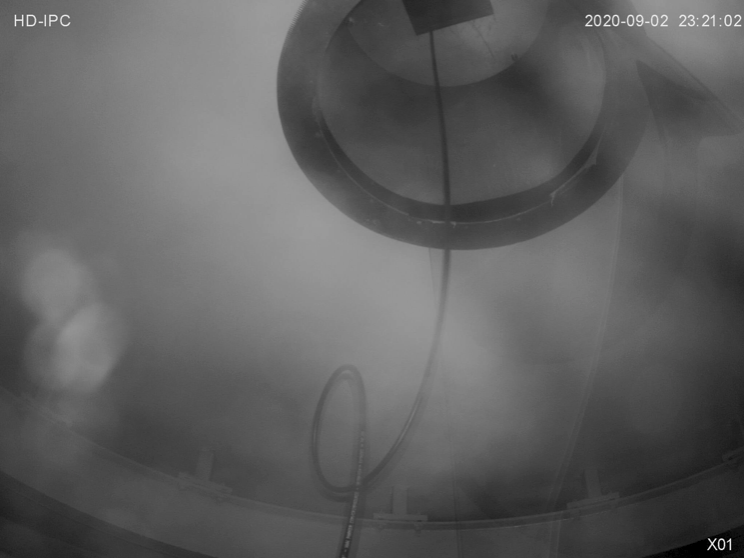}
\includegraphics[width=0.49\linewidth]{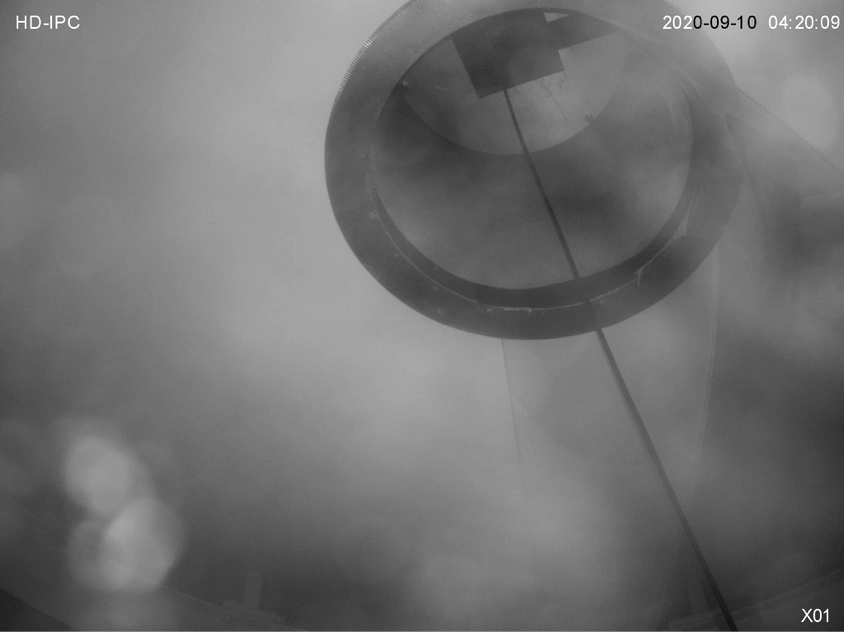}
\vspace{2mm}
\caption{\textbf{Surveillance images of the fiber conduit above the spectrograph enclosure inside the Pier Lab}. A webcam is monitoring the fiber to detect loops (like the one in the left hand side image). Telescope motion is stopped if a loop is being dragged up towards the roof of the Pier Lab before the fiber conduit can uncoil.}
\label{fig:fiber}
\end{figure}

For MAROON-X we have chosen a very stiff and crush-resistant conduit material with metal braided support and abrasion resistant outer cover. The conduit resists twisting which is important since motions of the telescope in azimuth and motion of the Cassegrain rotator impart twisting forces that are spread out by the conduit and ultimately translated into flips of the coiled portion of the fiber at the spectrograph end. We use a webcam to monitor the motion of the fiber conduit in the Pier Lab during slews of the telescope (see Fig.~\ref{fig:fiber}).

The fiber conduit is being dragged up and down during elevation motion of the telescope. Excess fiber conduit length is coiled up on top of the spectrograph enclosure when the telescope is not at minimum elevation but pointing near zenith. At the telescope end, the fiber conduit is guided by a fiberglass support rod at the bottom of the Ballast Weight Assembly (BWA) to form an arc (see top left hand side insert in Fig.~\ref{fig:overview}) with a large bend radius to minimize stress on the optical fibers. We have added a roller guide to the cover plate guiding the fiber conduit down to the Pier Lab (see right hand side insert in Fig.~\ref{fig:overview}), to minimize friction and abrasion at this junction. This setup is the same as for GRACES\cite{GRACES1,GRACES2}, except that the GRACES fiber conduit never reaches the Pier Lab but is routed towards the CFHT from the Service Room on top of the Pier Lab.

\subsection{Frontend}
\label{sec:frontend}
\begin{SCfigure}[][!t]
\centering
\includegraphics[width=0.65\textwidth,keepaspectratio]{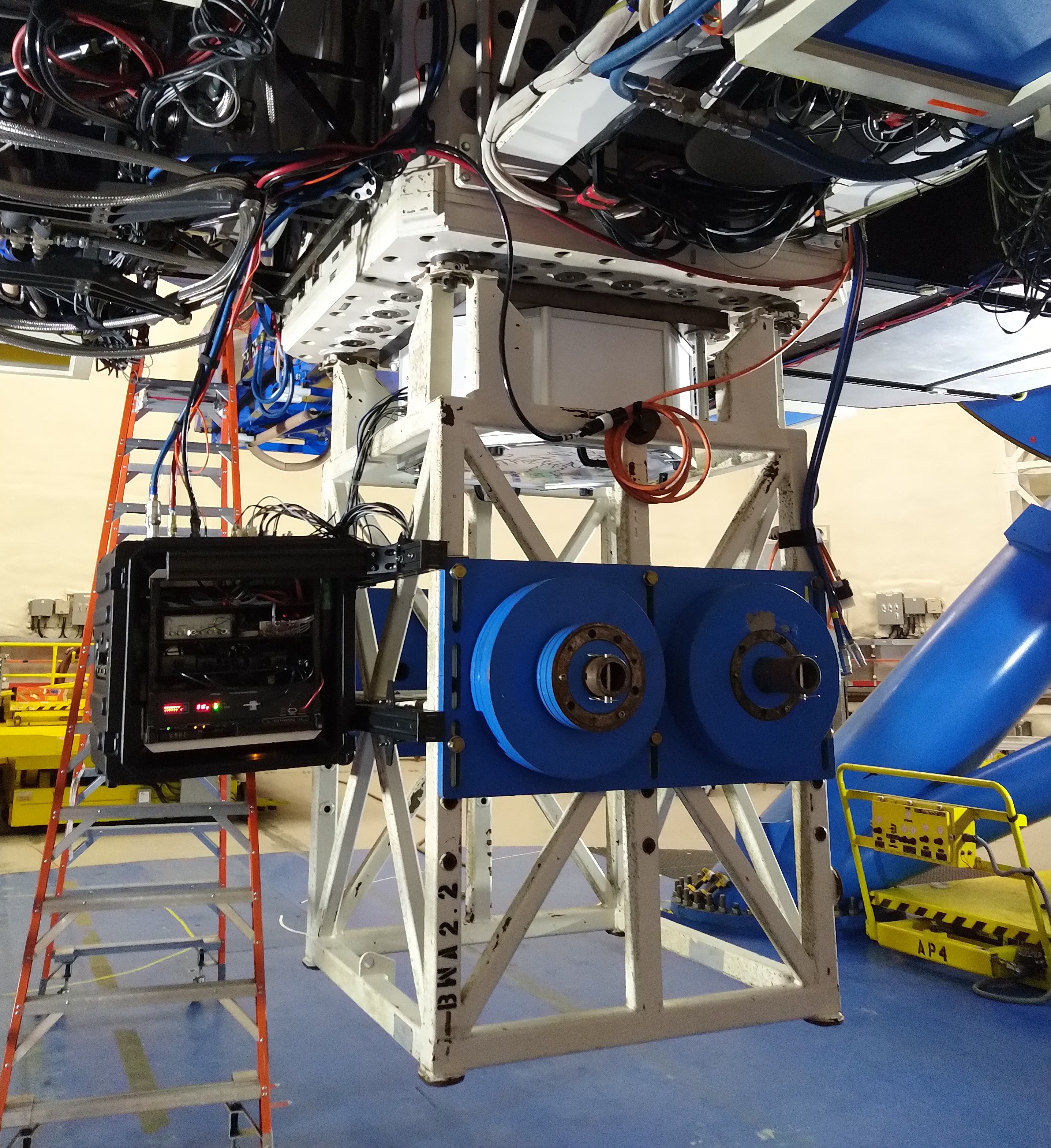}
\caption{\textbf{MAROON-X frontend mounted on the bottom port of Gemini North's ISS}. In this image the electronics rack has been opened for service and the fiber conduit is not yet attached.}
\label{fig:frontend}
\end{SCfigure}

The technical requirements and the opto-mechanical design of the telescope frontend have been presented in Ref.~\citenum{mx2018}. The frontend is seated inside a dedicated Ballast Weight Assembly (BWA). The electronics are mounted in a 10U \SI{19}{\inch} rack inside an insulated SKB shockmount roto case. The electronics rack is mounted on the side of the BWA in compliance with Gemini requirements (see Fig~\ref{fig:frontend}).

The BWA also serves as a rig to mount the frontend on the ISS and as a safe storage location when the frontend is not on the telescope. Since we share the bottom port of the ISS with Gemini facility instruments NIRI and NIFS, the mechanical design of the frontend was driven by the requirement of making the instrument swap as easy as possible. This includes a simple, repeatable, and safe way of installing and removing the fiber conduit from the frontend. As described in Ref.~\citenum{mx2018}, the fibers are embedded in a fused silica plate that is mounted in a ``plug" that can be inserted into the fiber focal plane of the frontend. The plug uses a three-point groove and ball bearing mechanical reference for repeatable seating and is rigidly held in place with a simple clamp mechanism.

We commissioned the frontend already in December 2018 to address potential issues ahead of the installation and commissioning of the spectrograph. We used a few hours of on sky time to validate the ADC and guiding performance. We show part of the frontend control interface in Sec.~\ref{sec:control_software} and discuss performance issues in Sec.~\ref{sec:performance}

\subsection{Spectrograph}
\label{sec:spectrograph}
The core spectrometer of MAROON-X is a dual-arm KiwiSpec R4-100\cite{barnes2012,gibson2012} with a number of design modifications to  move from the iodine cell method used in its MINERVA\cite{wilson2019} variant to a level of intrinsic stability that permits the use of the less restrictive and more efficient simultaneous calibration technique employed in high-performance instruments such as, e.g., HARPS\cite{HARPS} or ESPRESSO\cite{pepe2020}. The opto-mechanical design of our spectrograph as well as the characterization of individual optical components of the spectrometer have been reported at previous SPIE conferences\cite{mx2016,mx2018}. 

\begin{SCfigure}[][!b]
\centering
\includegraphics[width=0.65\textwidth,keepaspectratio]{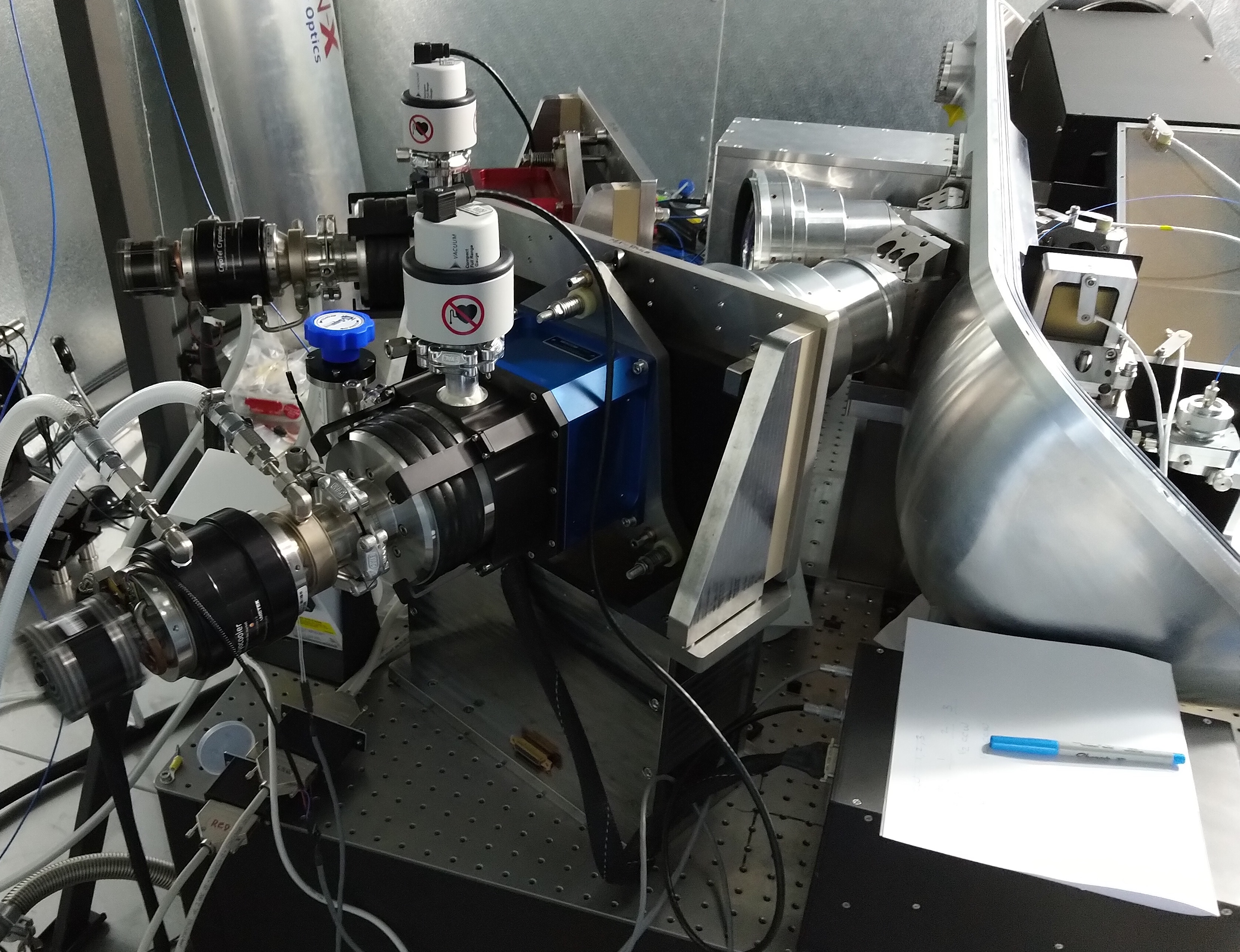}
\caption{\textbf{CCD detector systems of MAROON-X at Gemini Observatory} during the AIT phase. Two almost identical detector systems record the echelle spectra behind the red (NIR) arm (back) and the blue (vis) arm (front). One of the Archon CCD controllers is visible in the lower right. The CCD controllers were eventually mounted away from the main optical table on separate stands and enclosed in insulating boxes. Cryocoolers and CCD controllers are water-cooled. A thin Bonn shutter is placed right before each dewar window.}
\label{fig:ccd}
\end{SCfigure}

The two science grade detector systems and the red camera arm were delivered later than anticipated and due to Hofstadter's Law\footnote{Hofstadter's Law: It always takes longer than you expect, even when you take into account Hofstadter's Law.} only basic end-to-end testing of the complete spectrograph was possible before it was disassembled and shipped to the observatory.

\begin{figure}[!b]
\centering
\includegraphics[width=0.49\textwidth,keepaspectratio]{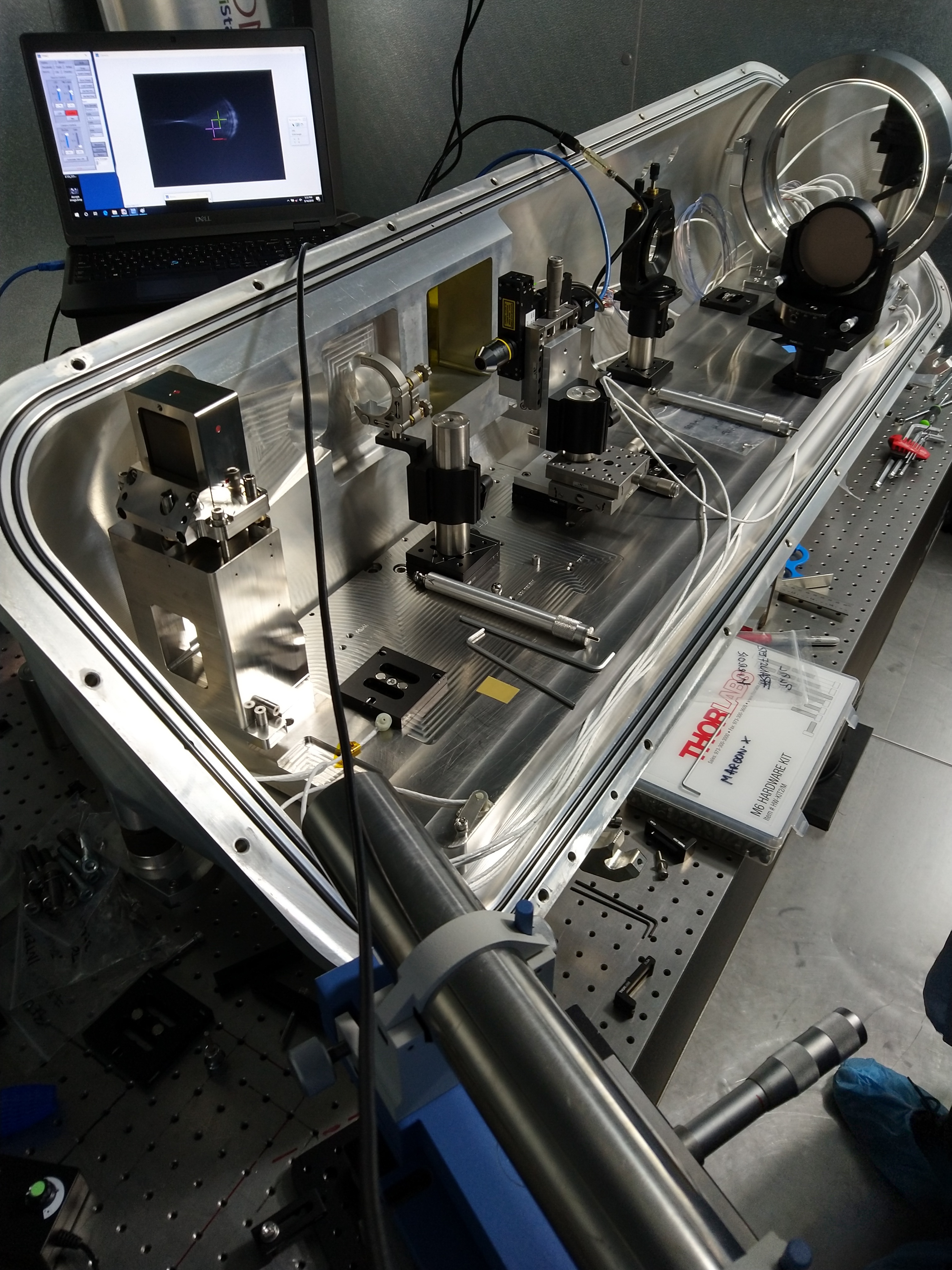}
\hspace{1mm}
\includegraphics[width=0.49\textwidth]{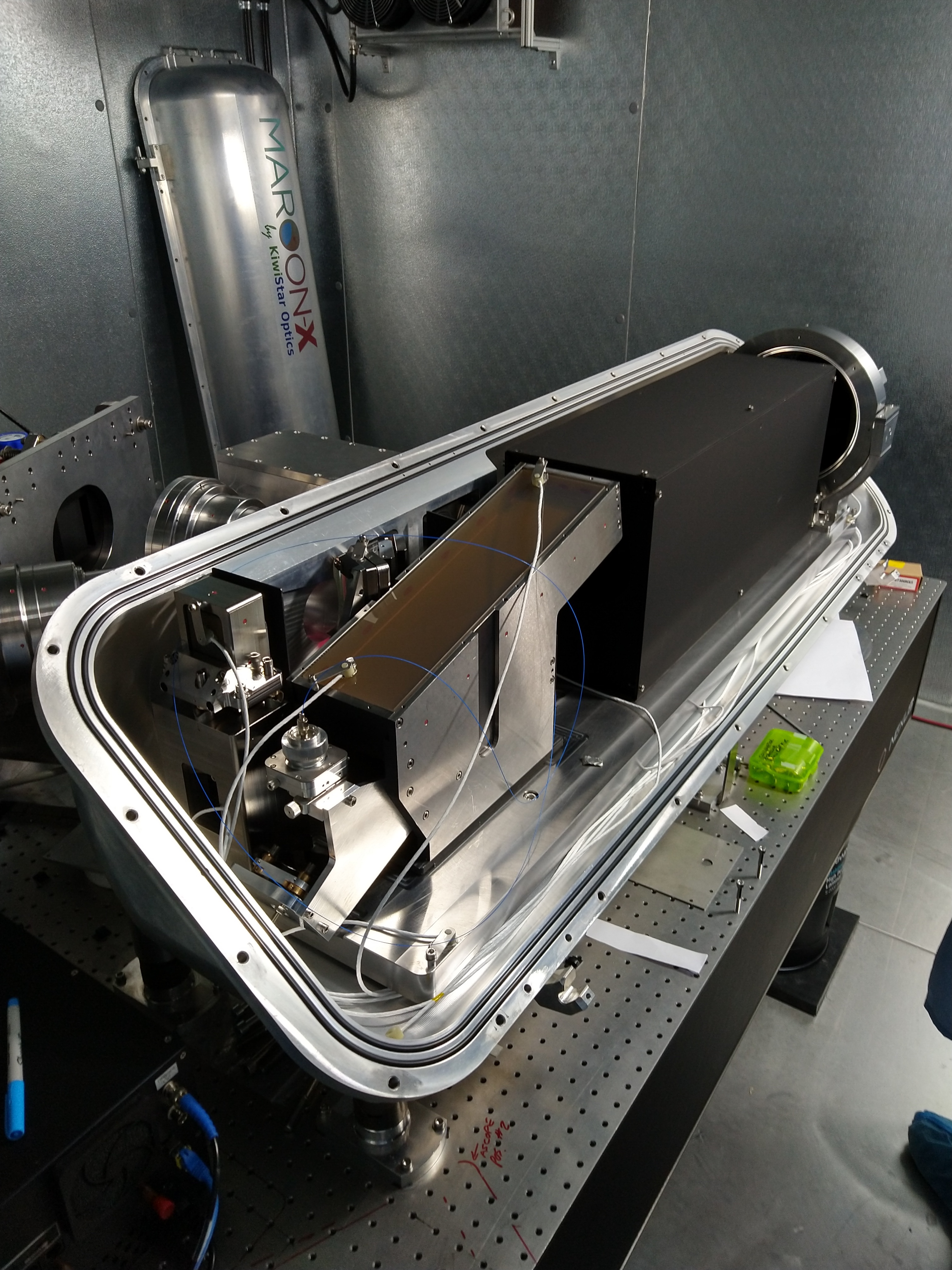}
\vspace{0mm}
\caption{\textbf{Optical alignment of MAROON-X at Gemini Observatory.} \textit{Left:} At early stages of the alignment procedure, with the main collimator (top right) and the pupil transfer mirror (left) on the optical bench. A point source microscope and an autocollimating alignment telescope are used in conjunction with mechanical fiducials to establish the optical axis and align components. \textit{Right:} After completion of the alignment, before closing the vacuum chamber. A baffle assembly is placed between the echelle grating (mounted face down) and the main collimator. The assembly at the back-end of the echelle grating mount is part of the exposure meter. PT100 temperature sensors are mounted to various components.}
\label{fig:alignment}
\end{figure}

Thanks to in-built opto-mechanical fiducials and a meticulous alignment plan\cite{Gers2014}, the spectrograph was assembled and aligned at Gemini North in under two weeks (see Fig.~\ref{fig:alignment}). Subsequently, the CCD detector systems were cooled down and focused (see Fig.~\ref{fig:ccd}).

While installing the spectrograph at Gemini, we implemented a small number of modifications to the original design. The most significant is the mechanical design and location of the pupil slicer. In order to match the modest 100\,mm beam size of MAROON-X to the large aperture of the Gemini telescope while at the same time maintaining an acceptable FOV and resolving power, we employ a pupil slicer. The pupil slicer reformats the pupil at the exit of the science fiber and projects the slices onto three rectangular fibers that are stacked to form a pseudo-slit at the entrance of the spectrograph. The pupil slicer is based on micro lenses and incorporates a double scrambler. Details of the optical design and prototyping were reported in Ref.~\citenum{slicer2016}. 

\begin{SCfigure}[][!t]
\includegraphics[width=0.6\linewidth]{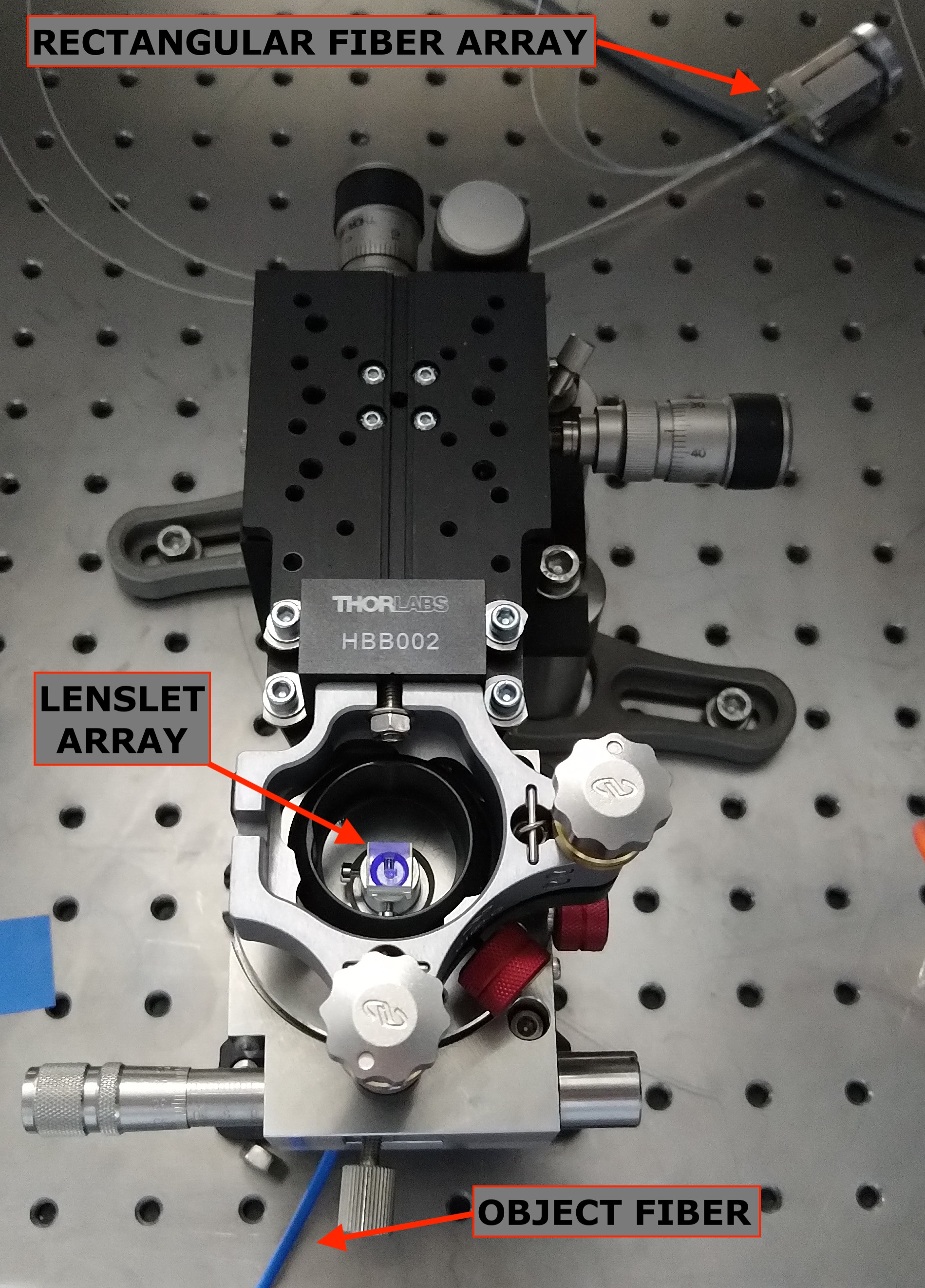}
\caption{\textbf{Final mechanical setup of the pupil slicer for MAROON-X}. A pupil image is formed from the light exiting the \SI{100}{\micro\meter} octagonal science fiber by a custom miniature achromat (not visible). The pupil image is projected onto the underside of a double-sided lenslet array mounted in a rotation stage. The fiber and achromat are mounted in a flexure stage such that the XY position of the pupil image can be adjusted relative to the lenslet array. A linear array of 5 rectangular fibers is mounted in a jig (seen lying on the table in the top right corner of the image). When the jig is placed on top of the lenslet array, the central three rectangular fibers receive the three slices of the pupil image formed by the lenslet array. The jig is held in a gimbal mount which itself is mounted on a XYZ flexure stage, allowing for focus (Z), lateral (XY) and tip-tilt ($\Theta_X,\Theta_Y$) adjustment. This setup is mounted on the optical bench of the spectrograph, benefiting from its mechanical and thermal stability.}
\label{fig:slicer}
\end{SCfigure}

The mechanical design concept of the pupil slicer called for an installation inside a small rectangular extension unit to the main vacuum chamber. We designed custom vacuum-compliant mechanics for the pupil slicer accordingly. While moving towards integration in the lab, we realized that the space constraints imposed by the limited size of the extension unit imposed a significant risk of damaging fibers while aligning the pupil slicer at the observatory. We thus decided to set up the pupil slicer outside the vacuum chamber on the main optical table of the spectrograph. COTS components were ultimately used to construct the mechanics of the pupil slicer, allowing for easy and robust alignment (see Fig.~\ref{fig:slicer}). 

An additional minor change to the original design concerns the fiber agitator. Both the original design of the pupil slicer as well as its current implementation allow for the agitation of the rectangular fibers directly before entering the spectrograph in order to suppress model noise. The fibers form a double loop and are hanging suspended from a ``bridge''. In the lab we saw that slight stretching motions of this loop highly randomizes modal noise patterns seen in coherent light sources. We were thus planning to use a voice coil to impart randomized electro-magnetic forces to a magnet attached to the bottom of the fiber loop. However, we found no sign of modal noise induced RV variations in the stellar light. Since we also employ additional COTS fiber modal noise scramblers for the FP etalon comb calibrator, we decided to not use the agitator until we have a clear indication that residual modal noise is limiting our performance.

\subsection{Exposure meters}
The primary exposure meter uses a COTS OAP mirror (76\,mm diameter, 152\,mm EFL) to collect a fraction of the 0th order of the echelle grating and focus a section of the entrance slit image containing the three central object fibers onto a \SI{300}{\micrometer} high-NA fiber. The fiber is then routed out of the spectrometer and a set of lenses re-images the fiber onto a cooled, back-thinned CCD. The detector (Hamamatsu S7031-1007) has excellent efficiency (peak QE $>$90\% at 650\,nm) and acceptable dark current ($\sim$1.5\,e$^-$\,pixel$^{-1}$\,sec$^{-1}$) and read noise ($\sim$ 15\,e$^-$pixel$^{-1}$). However, in practice we find that without binning the \SI{24}{\micrometer} CCD pixels, the sensitivity of the exposure meter is very poor, even for only moderately faint sources (R$>$12\,mag). 

We thus use a second channel of the exposure meter that was originally intended for long-term monitoring of the focal-ratio-degradation (FRD) effects in the science fiber. As explained above, three lenses in the pupil slicer lenslet array slice the pupil image and re-image the three slices onto three rectangular fibers that form the central three sections of the spectrometer's pseudo slit. FRD increases the size of the pupil image and additional lenslets in the array focus a part of this ``FRD spill-over" light into two extra rectangular fibers. One of these fibers is then re-imaged onto the exposure meter CCD. The signal is read out simultaneously with that of the primary channel. Due to the FRD effects we discuss in Sec.~\ref{sec:performance}, and the much smaller surface area of this fiber, we achieve a much better signal-to-noise ratio on this channel, allowing exposure times of \SI{<2}{\sec} for sources approaching R$\sim$16\,mag. We have verified on bright stars that there are no major differences between the flux received by the FRD channel and the main channel, indicating that FRD effects are stable in time over at least a single exposure. The FRD channel can thus be used in lieu of the main channel for faint sources.

\subsection{Calibration System}
\label{sec:calibration}
A number of light sources are available for calibration. We use a tungsten-halogen light source with a color balancing filter for the flatfield illumination, and a Photron ThAr lamp and a custom Fabry-P\'erot (FP) etalon\cite{stuermer2016} for wavelength calibration and RV drift correction. All light sources are fiber coupled. We use COTS multimode fiber modal noise scramblers to suppress the modal noise in the FP etalon, which is illuminated by a NKT \textit{SuperK compact} continuum laser. COTS fiber switchers allow fast and repeatable selection of light sources which can be sent simultaneously and independently to the frontend unit for calibration of the science and sky fiber as well as to the simultaneous calibration fiber for RV drift monitoring. The feed to the simultaneous calibration fiber has a continuous neutral density filter wheel placed in a non-collimated light path to automatically adjust the flux rate depending on the exposure time without introducing any fringing.

The FP etalon is referenced against a set of hyperfine transitions of rubidium to compensate for long term drifts due to temperature fluctuations or aging of the Zerodur spacer material\cite{stuermer2017}. The rubidium referencing is the last major subsystem that was not ready for the commissioning run in December 2019 and remained unused due to the access restriction to the observatory in 2020. We have thus been using the ThAr source to check for long-term drifts of the FP etalon. We can not detect drifts exceeding the precision achievable with our ThAr source ($\simeq$ 70\,cm\,s$^{-1}$) over two weeks, the typical duration of an observing run. The impurities from ThO in our ThAr lamps\cite{dirt} limit the precision achievable with this technique and re-enforce our desire to re-commissioning the rubidium referencing during the first intervention in 2021.

\subsection{Environmental control}
\label{sec:envcontrol}
We actively control the air temperature inside the spectrograph's walk-in enclosure with a Huber unitstat model 405w and two Lytron dual-fan heat exchangers. A PT100 sensor measures the temperature of the air stream exiting one of the two heat exchangers and a PID loop in the unistat controls the cooling and heating power. We achieve a long-term stability of the air temperature of about 20\,mK P-V, measured at a separate sensor in the air stream of the second heat exchanger. 
\begin{figure}[b!]
\centering
\includegraphics[trim={0cm 1cm 0cm 0cm},width=0.99\textwidth,keepaspectratio]{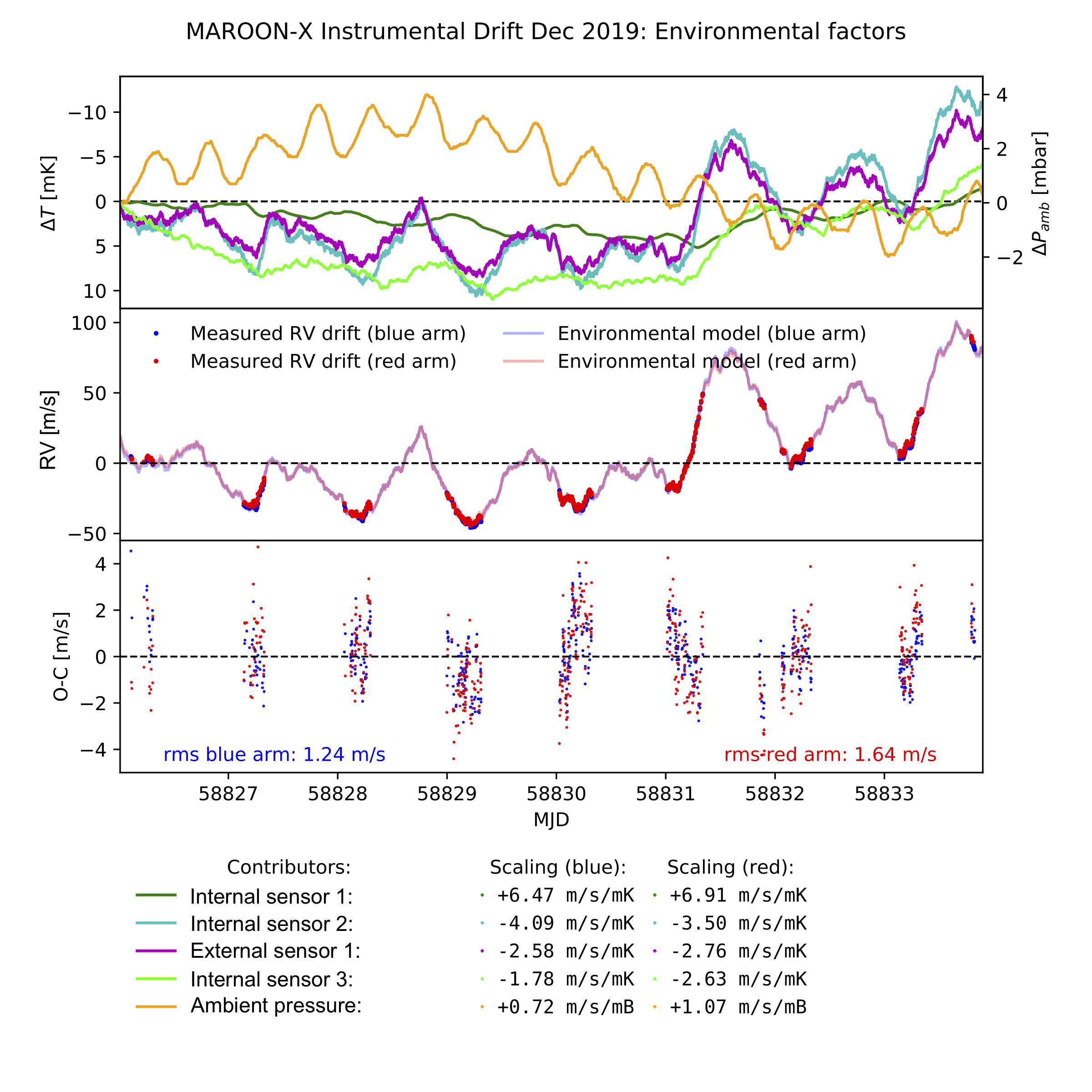}
\vspace{2mm}
\caption{\textbf{Absolute RV drift of MAROON-X and comparison to temperature measurements} during the commissioning phase in December 2019. {\it Top:} Measurements from four of the 16 temperature sensors on the spectrograph as well as the  ambient pressure as a function of time. {\it Center:} Average absolute RV zeropoint relative to the first recorded observation. Datapoints derived from FP etalon spectra taken through the science fiber of MAROON-X are marked as red and blue dots for each of the two arms, respectively. Red and blue lines show a composite model fitted to those points based on the four temperature sensor data and ambient pressure shown in the top panel. The coefficients of the model are listed in the legend. {\it Bottom:} Residuals between the observed RV zeropoint drift and the composite model.}
\label{fig:rvstability}
\end{figure}

Short-term temperature fluctuations in the Pier Lab are greatly dampened by the enclosure walls but daily and seasonal drifts still have a tendency of reaching the spectrograph through conductive paths, despite extensive efforts to insulate the pneumatic legs of the outer optical table. As a result, the spectrograph temperature underwent seasonal drifts of up to 200\,mK over timescales of weeks to months and saw diurnal variations of 10--15\,mK P-V. This was particularly pronounced when the observatory was in shutdown and the dome was not pre-cooled for the night, causing larger differences in the day- and nighttime temperatures inside the building. Diurnal variations are further dampened by a factor of $\sim$5 before reaching critical components inside the vacuum chamber, like the echelle grating or the internal optical table. Seasonal variation fully impact all components inside the vacuum tank. 

These temperature fluctuations have a profound impact on the instrument's absolute RV stability. In Fig.~\ref{fig:rvstability} we show the RV zeropoint drift of MAROON-X as a function of time during the commissioning run in December 2019. The instrument drifted almost 140\,m\,s$^{-1}$ over one week, measured as the average drift of all etalon lines in a calibration exposure in respect to the first exposure. Since data were taken only over a few hours at the beginning of each night and the simultaneous calibration fiber wasn't operational, we only see a small part of the diurnal drift that the instrument underwent. 

With data from only four temperature sensors and the ambient pressure value we can form a simplistic model that reproduces the RV drift to better than 2\,m\,s$^{-1}$ rms. In the model we scale the sensor values and allow for a phase shift due to different levels of insulation and dampening experienced by each sensor. The limiting factor for the model is the precision of the sensor data. With scaling factors ranging from 1.8--7\,m\,s$^{-1}$\,mK$^{-1}$, we need temperature data with a precision of well under 1\,mK to fit the data at the m\,s$^{-1}$ level. This can be achieved by averaging over the recorded raw sensor data but actual temperature fluctuations occur on timescales of 10--15\,min, limiting the ability to bin longer than that.

With the simultaneous calibration fiber working in 2020, we have continuous measurements of the instrumental RV zeropoint over much longer timescales and with regular cadence. We are in the process of analysing these data to further guide the data analysis and our understanding of the impacat of environmental factors on the RV stability. The measured scaling factors indicate a much higher temperature sensitivity of the instrument than the initial thermo-mechanical design suggested. While our etalon-based drift measurements can compensate for the majority of the absolute drift (see Sec.~\ref{sec:rvperformance}), further improvement to the absolute stability of the instrument can only be beneficial.

As a first measure, we have stopped the forced venting of the Pier Lab at the end of July of 2020 and found that seasonal changes are now greatly suppressed and the diurnal temperature variations all but vanished (see, e.g., Fig.~\ref{fig:grafana}). Active temperature control of the Pier Lab is on the list of interventions planned for 2021 to further improve the temperature stability of the spectrograph. In addition we plan to tie the control loop of the unistat to the average spectrometer temperature instead of the air temperate inside the enclosure. 

\section{Software}
\subsection{Instrument control}
\label{sec:control_software}
The instrument control software for MAROON-X is based on a client-server architecture using \texttt{gRPC}\footnote{\texttt{https://grpc.io/}}, a modern open source high performance RPC framework. Server and clients are written in \texttt{python3}. \texttt{Redis}\footnote{\texttt{https://redis.io/}}, an open-source in-memory data structure store (a key–value database) is used to facilitate the distribution of 
instrument status information. \texttt{Redis} clients grab a snapshot of the database for inclusion into data headers and for permanent storage in \texttt{InfluxDB}\footnote{\texttt{https://www.influxdata.com/products/influxdb/}}, an open-source time series database. 

\begin{figure}[t!]
\centering
\includegraphics[width=0.95\textwidth,keepaspectratio]{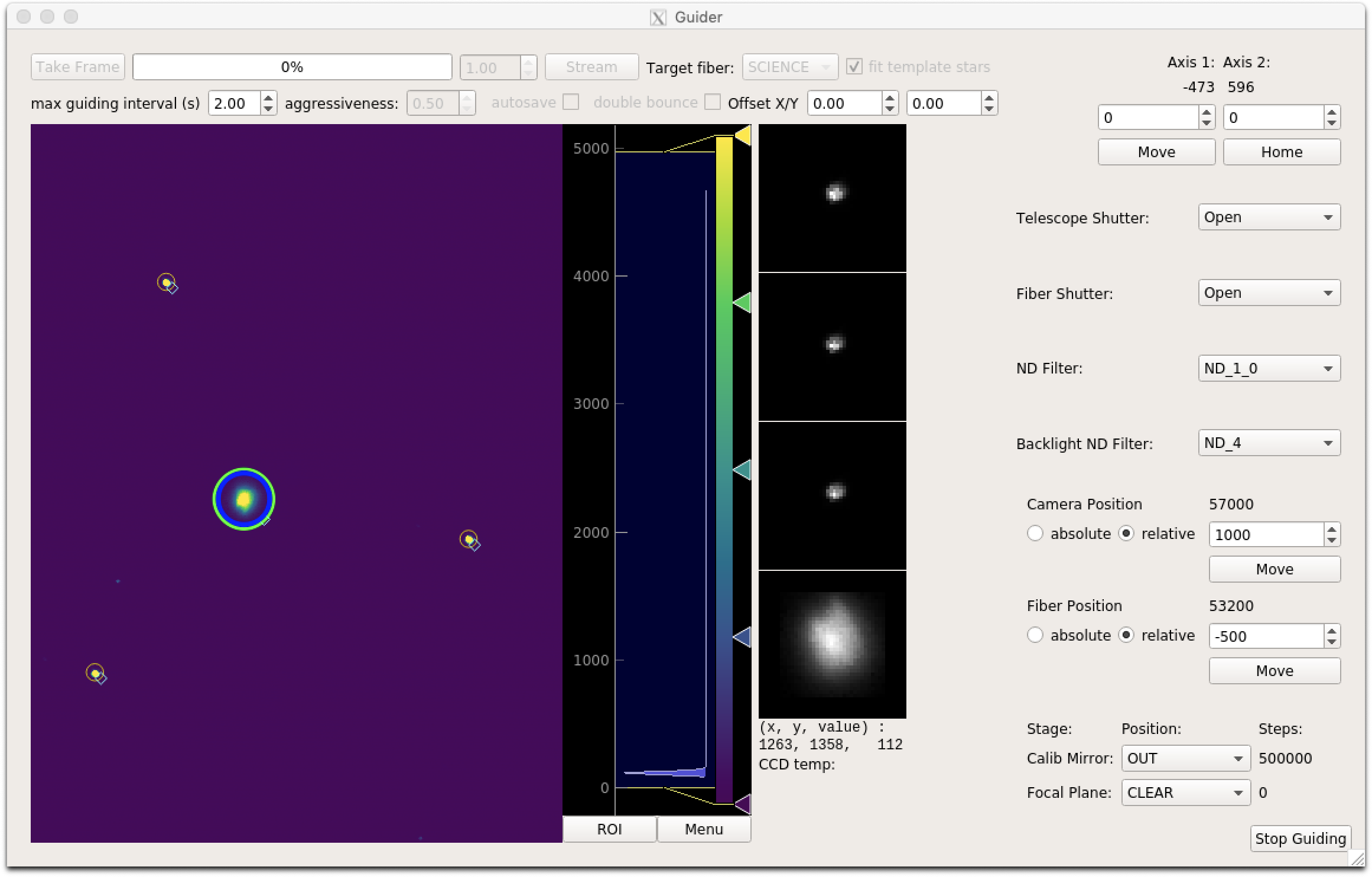}
\vspace{2mm}
\caption{\textbf{MAROON-X guider widget} during a typical observation. The three sources forming a triangle are back-illuminated single-mode fibers that are used to triangulate the instantaneous position of the science fiber, marked by the blue circle. The centroid of the stellar image (green circle) is measured and the guiding algorithm is sending commands to a tip-tilt stage to keep the star centered on the fiber. The guiding algorithm allows for independent control of the guider exposure time and guide loop update rate. A user selectable aggressiveness setting controls the fraction of the error signal that is send to the tip-tilt mirror. This widget also includes controls for other components in the frontend, such as shutters, focus stages, neutral density filter wheels, and calibration stages.}
\label{fig:guider}
\end{figure}

The graphical user interface (GUI) is comprised of a suite of \texttt{Qt}\footnote{\texttt{https://www.qt.io/}} widgets implemented in \texttt{python3} and executed in a VNC environment. In Fig.~\ref{fig:guider} we show the guider widget as an example. This widget controls the sCMOS guide camera, the tip-tilt mirror, two focus stages, two neutral density filter wheels, two shutters, and two calibration stages through their respective clients. It also includes a state machine that controls the actual guiding algorithm that keeps the centroid of the star on top of the science fiber. 

The VNC environment allows for seamless remote observations and easy handovers between observers.

\subsection{Telemetry}
A suite of PT100 sensors read out by CryoCon 18i monitors are used to record air temperatures in the Pier Lab and in the spectrograph enclosure, as well as  surface temperatures of critical components inside and outside of the spectrograph's vacuum tank. The data are stored in the \texttt{InfluxDB} database.

We also store health and performance data from a number of components such as the Archon CCD controllers, SunPower cryocoolers, vacuum gauges, and vibration sensors. When in operation, the rubidium tracking system of the FP etalon also stores fit parameters and health data in the database.

\begin{figure}[h!]
\centering
\includegraphics[width=0.99\textwidth,keepaspectratio]{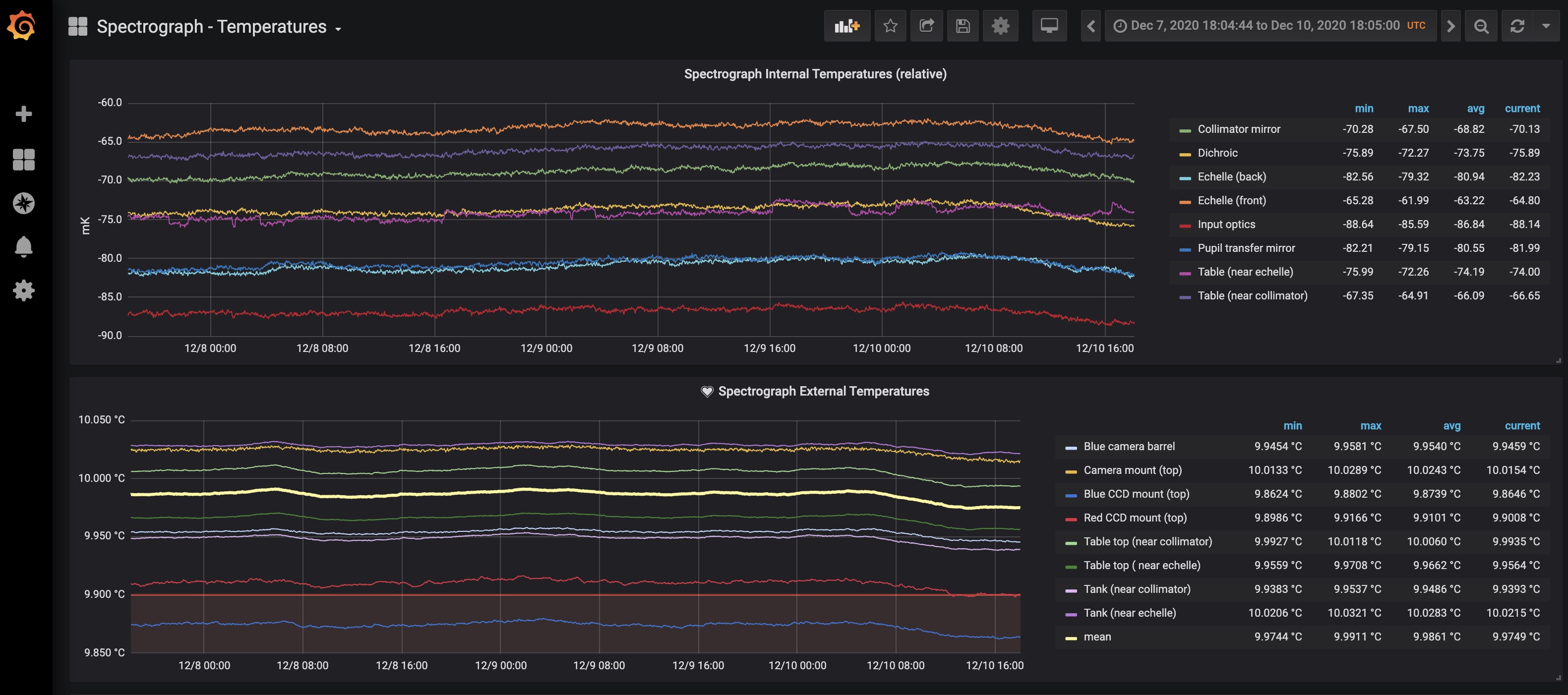}
\vspace{2mm}
\caption{\textbf{Snapshot of a Grafana dashboard showing temperature sensor data inside and outside the spectrograph's vacuum tank}. Alarms are automatically triggered if lower or upper limits defined for selected sensors (red horizontal lines in the lower panel) are exceed.}
\label{fig:grafana}
\end{figure}

For visualisation and monitoring, we use \texttt{Grafana}\footnote{\texttt{https://grafana.com/}}, a multi-platform open-source analytics and interactive visualization web application. We have email alerts set up in \texttt{Grafana} to notify us when sensor readings fall outside of acceptable ranges, indicating some form of malfunction or drastic change in environmental conditions. Fig.~\ref{fig:grafana} shows a snapshot of a dashboard monitoring temperature sensor data inside and outside the spectrograph's vacuum tank.

\subsection{Data reduction and RV analysis}
\label{sec:analysis_software}
Standard data reduction procedures, including bias and background subtraction, order tracing, and optimal flux extraction are implemented in a suite of custom \texttt{python3} routines and scripts. This software set also extracts the emission lines from the FP etalon comb present in either the science fibers in a calibration frame or the simultaneous calibration fiber in the science spectra and fits them with a simple but well matched analytic profile. The positions of the etalon lines are used to compute the wavelength solution and the drift correction by fitting cubic smoothing splines to the etalon line positions for each fiber and each order of the spectrum. In the future we plan to implement all data reduction routines in \texttt{Dragons}, Gemini's new python-based data reduction platform\cite{dragons}. 

For the RV analysis, we have two working radial velocity pipelines: one based on \texttt{SERVAL}\cite{serval}, and another based on \texttt{wobble} \cite{wobble}. Both packages are freely available\footnote{\texttt{https://github.com/mzechmeister/serval}, and \texttt{https://github.com/benmontet/wobble}} and work with MAROON-X data after only modest modifications.

\texttt{SERVAL} employs a template-matching algorithm that has been shown to outperform the classical cross-correlation function (CCF) technique on M dwarfs. \texttt{SERVAL} was originally developed for the CARMENES\cite{CARMENES} project. It has been thoroughly vetted and was used for a number of scientific publications based on CARMENES data. In addition to RVs, \texttt{SERVAL} also computes activity indicators, such as the chromatic index, differential line width, and indices for prominent lines such as H$\alpha$, Na D, or the Ca IRT. These indices allow a first assessment of the impact of stellar activity on the RV data and help diagnose problems with the masking of telluric lines. \texttt{SERVAL} is currently our default RV analysis code. 

\texttt{wobble} is a data-driven analysis technique for time-series stellar spectra. It uses \texttt{TensorFlow} to fit a linear model in $\log{F}$ to solve for precise RVs. It infers the underlying stellar and telluric spectra without morphological priors. \texttt{wobble} has been vetted against the HARPS DRS. Initial tests using \texttt{wobble} for MAROON-X spectra have confirmed the RV results obtained with \texttt{SERVAL}, an important cross-validation. 

\section{Performance}
\label{sec:performance}
\subsection{Throughput}
\label{sec:throughput}

Efficiencies were measured for all individual components of the spectrometer as well as for the as-built system in the lab. KiwiStar reported a peak throughput of the spectrometer from the pseudo-slit to the focal plane of over 60\% in both arms\cite{mx2016}. Based on the measured throughout of the frontend of $\simeq$88\% from 500--900\,nm, estimates of the Gemini telescope efficiency (90\%), geometric coupling losses at the fiber feed (FOV = \SI{0.77}{\inch}) depending on seeing conditions (52\% at IQ75, i.e. \SI{0.7}{\inch} FWHM), an estimated 75\% efficiency of the pupil slicer, and 90\% peak QE of the CCDs, we were expecting a peak throughput of up to 15\% for the complete system.

\begin{figure}[t!]
\centering
\includegraphics[width=0.99\textwidth,keepaspectratio]{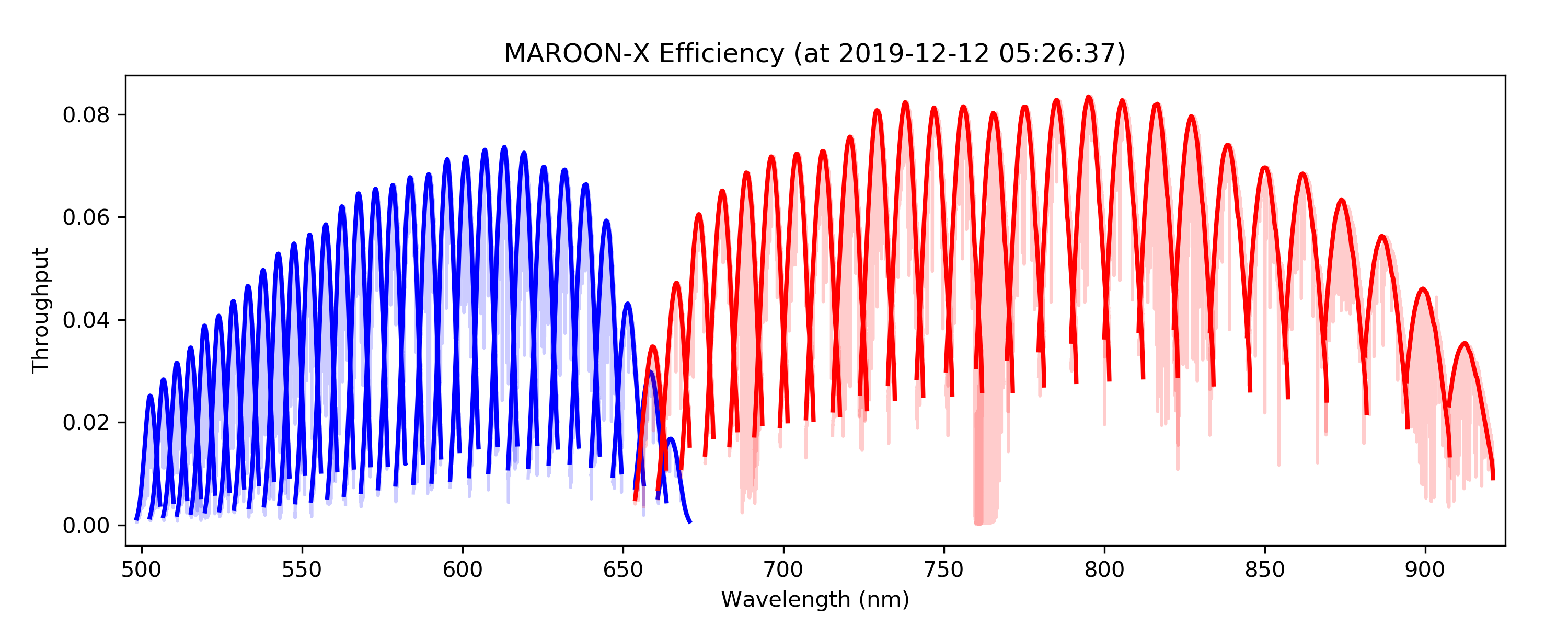}
\caption{\textbf{System throughput measured under median seeing conditions} (DIMM seeing $\simeq$\SI{0.6}{\inch}) during the commissioning run in December 2019}
\label{fig:throughput}
\end{figure}

Our on-sky measurements of the throughput fall significantly short of this number. For median seeing conditions of \SI{0.6}{\inch} we measure a peak system throughput of 8\% in the red (NIR) channel and 7\% in the blue (VIS) channel, see Fig.~\ref{fig:throughput}. While this is at par or even exceeds other RV spectrographs, both operational (HARPS-N: 8\%\footnote{\texttt{http://www.tng.iac.es/instruments/harps/}}, HARPS: 6\%\cite{HARPS_UM}, CARMENES VIS: 7\%\cite{CARMENES_VIS}),  and under construction (NEID HR mode: 6\%\cite{NEID_website}, KPF: 8\%\cite{KPF2016}), we aim to improve this number in the future.

So far we have identified two major causes of unaccounted throughput losses: We found a mismatch between the telescope pupil and the pupil formed from the back-illuminated science fiber in the telescope frontend. This mismatch, which is basically a misalignment of the chief ray angle at the fiber entrance, manifests as FRD losses at the spectrograph\cite{Avila98} (or rather the pupil slicer in our case). We confirmed the FRD by imaging the output pupil of the science fiber, which no longer shows the central obscuration of the secondary mirror, a clear sign of significant FRD\cite{adam2016}. As a consequence, the central slice of the pupil slicer contains more flux than the outer two slices. This geometric feature would have been balanced out if the central slice would contain the obscuration from the telescope's secondary mirror.

We can exclude the fiber link itself as a significant contributor to the FRD, as we had tested the fiber link in the lab before installing it at Gemini and and found no reason for concern. To exclude any potential damage to the fiber link during installation, we removed the fiber link after our first-light tests, re-tested it in our lab in Chicago, and re-installed it for the science verification run in December 2019. We found no change in the FRD behavior before and after use on Gemini. Due to access restriction to the observatory imposed by the COVID-19 pandemic, we were not able to address the frontend alignment issue in 2020. Fixing this issue is on our priority list for planned instrument interventions in 2021.

The second contributor to the throughput losses is an unexpected image quality 
issue we encountered at the Gemini North Telescope. During the first-light, commissioning, and subsequent science campaigns, we find a much higher complexity in effective seeing conditions than what is normally expressed with just a simple FWHM number. Of specific concern is an excessive jitter term that at times becomes a strong component of the effective (long-time integrated) FWHM, compared to the FWHM in individual guider images with \SIrange[]{1}{2}{\second} exposure times. The tip-tilt guiding system in our frontend is not designed to exceed a guide rate of about \SI{1}{\hertz} as its main function is to spead up acquisition, correct beam skews from the ADC, and compensate for flexure between the telescope guider and our fiber focal plane. This jitter term thus remains uncorrected and can contribute significantly to the coupling losses. We have encountered periods of good seeing (\SIrange[]{0.4}{0.5}{\inch}) with high stability where our throughput peaks at 11.5\%, while similar conditions with low stability (i.e. significant jitter) can bring the throughput down below the 8\% found on average. 

At this point, we can exclude instrumental seeing or issues with the frontend optics after successfully imaging and guiding on a spatially unresolved calibration target provided by the Altair AO system at Gemini. We are in the process of comparing recorded guider images with DIMM seeing values as well as data from the Peripheral Wavefront Sensors (PWFS) of the Gemini Telescope to better understand these performance issues.

\subsection{Sensitivity}
\label{sec:sensitivity}
Owing to the large collecting area of the 8m Gemini North Telescope, and the instrument's red wavelength coverage, MAROON-X has an unprecedented sensitivity for faint M dwarfs. With the current system throughput we can achieve a peak SNR of $\sim$90 (at 820\,nm) per pixel in 30\,min for late M dwarfs as faint as V$\sim$16.5\,mag under median seeing conditions. 

\begin{figure}[b!]
\centering
\includegraphics[trim={0cm 0cm 0cm 0cm}, width=0.99\textwidth,keepaspectratio]{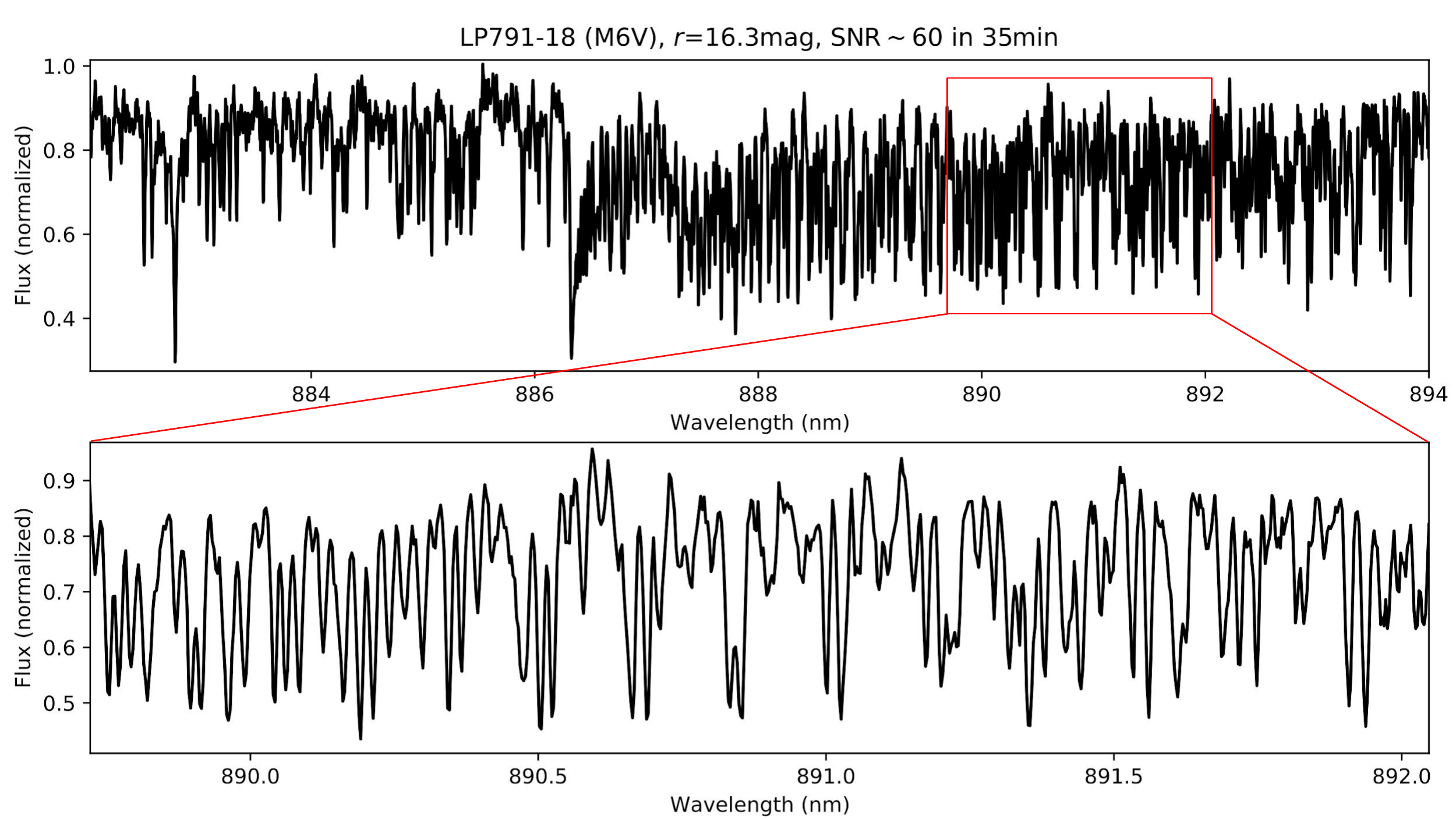}
\vspace{2mm}
\caption{\textbf{Spectrum of a M6V dwarf of $r$=16.3\,mag}. Observations taken under mediocre seeing conditions (DIMM $\simeq$\SIrange{0.7}{1.0}{\inch}) reached a peak SNR of $\sim$60 per pixel in 35\,min. The top plot shows echelle order 67, the bottom plot a cutout illustrating the density and depth of stellar lines. This specific echelle order is almost completely free of telluric lines.}
\label{fig:reach}
\end{figure}

M dwarfs provide a high level of intrinsic RV information due to their extreme spectral richness and observations with a moderate SNR yield useful RV precision. As an example we show a section of the spectrum of LP791-18, a faint M6V star with $r$=16.3\,mag in Fig.\ref{fig:reach}. Observations under worse-than-average seeing conditions yield a peak SNR of $\sim$60 per pixel in 35\,min. The intrinsic RV content of this exposure is 1.6\,m\,s$^{-1}$ for the red arm of MAROON-X.

\subsection{Resolving Power}
\label{sec:resolution}
We use the FP etalon comb lines to measure the resolving power of MAROON-X. 
The etalon lines have an intrinsic line width of \SI{340}{\mega\hertz} and are completely unresolved at R\,=\,80,000 (1/10--1/20 of a nominal resolution element), revealing the spectrograph's line-spread function (LSF). Our data reduction routines includes a step that fits an analytical profile constructed from a top hat (representing the slit geometry) and two Gaussians (approximating asymmetric aberrations) to each of the several ten thousand comb lines in each calibration frame. This model fits our LSF extremely well and can be used to precisely determine its effective FWHM.

\begin{figure}[t!]
\centering
\includegraphics[width=0.99\textwidth,keepaspectratio]{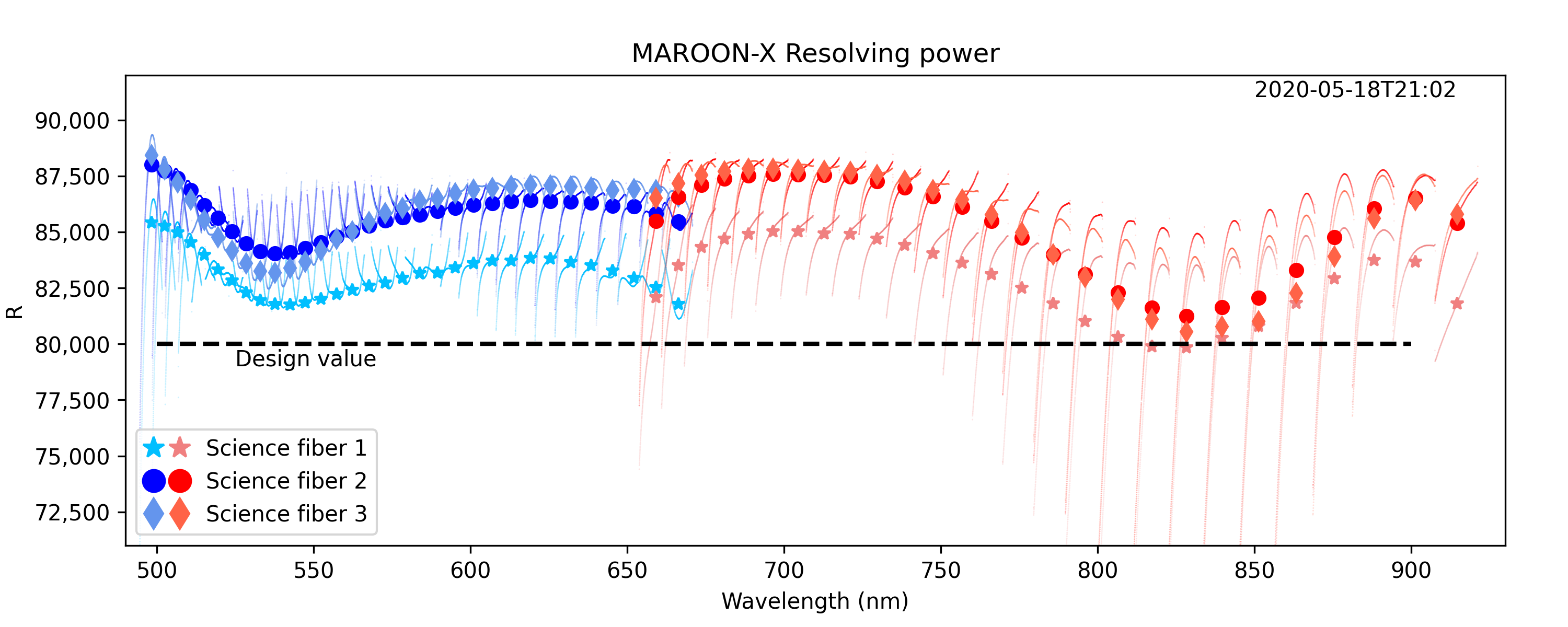}
\caption{\textbf{MAROON-X resolving power} for the three science fibers in the slit. Thin points show the raw values from each of the roughly 500 etalon lines per spectral order. Solid points mark the average value within the FSR of each order. The variations are entirely driven by the complex aberration pattern of the MAROON-X camera optics.}
\label{fig:resolution}
\end{figure}

Using the FWHM of the LSF as the $\Delta\lambda$ to determine the resolving power as $R=\lambda / \Delta\lambda$, we find the values presented in Fig.~\ref{fig:resolution}. On average, the resolving power exceeds the design value of R\,=\,80,000 and the mean over the complete spectrum is closer to R\,$\simeq$85,000. The variations are entirely driven by the complex aberration pattern of the MAROON-X camera optics. 

If desirable, one could balance the impact of the aberrations in the red camera arm by slightly adjusting the focus and tip-tilt of the detector system, but this would significantly impact the RV zero point before and after such an intervention.

\subsection{On-sky RV performance}
\label{sec:rvperformance}
The end-to-end RV performance of a spectrograph is the result of a large number of intertwined error sources\cite{Halverson2016}. A short list of critical performance figures would include the scrambling gain of the fibers feeding the spectrograph, the underlying stability of the wavelength reference source (ThAr, LFC, FP etalon) and a minimum of intrinsic stability that permits to translate drift measurements taken either simultaneously with each exposure or bracketing the science exposures to the stellar spectrum. 

We have validated the stability of our FP etalon calibrator during lab tests in 2016 and 2017\cite{stuermer2016,stuermer2017} and found that we can trace the zero point of the etalon to better than 10\,cm\,s$^{-1}$ at 780\,nm. We have also measured the differential drift between the science and calibration fibers to be under 20\,cm\,s$^{-1}$ with an engineering grade CCD that covered only 1/4 of the spectrum of MAROON-X's blue arm. We have also measured the scrambling gain of our octagonal science fiber\cite{adam2016}, but have yet to analyze end-to-end scrambling gain measurements that include the pupil slicer and double scrambler as well as the rectangular fibers that form the spectrograph's pseudo slit. 

\begin{figure}[t!]
\centering
\includegraphics[width=0.99\textwidth,keepaspectratio]{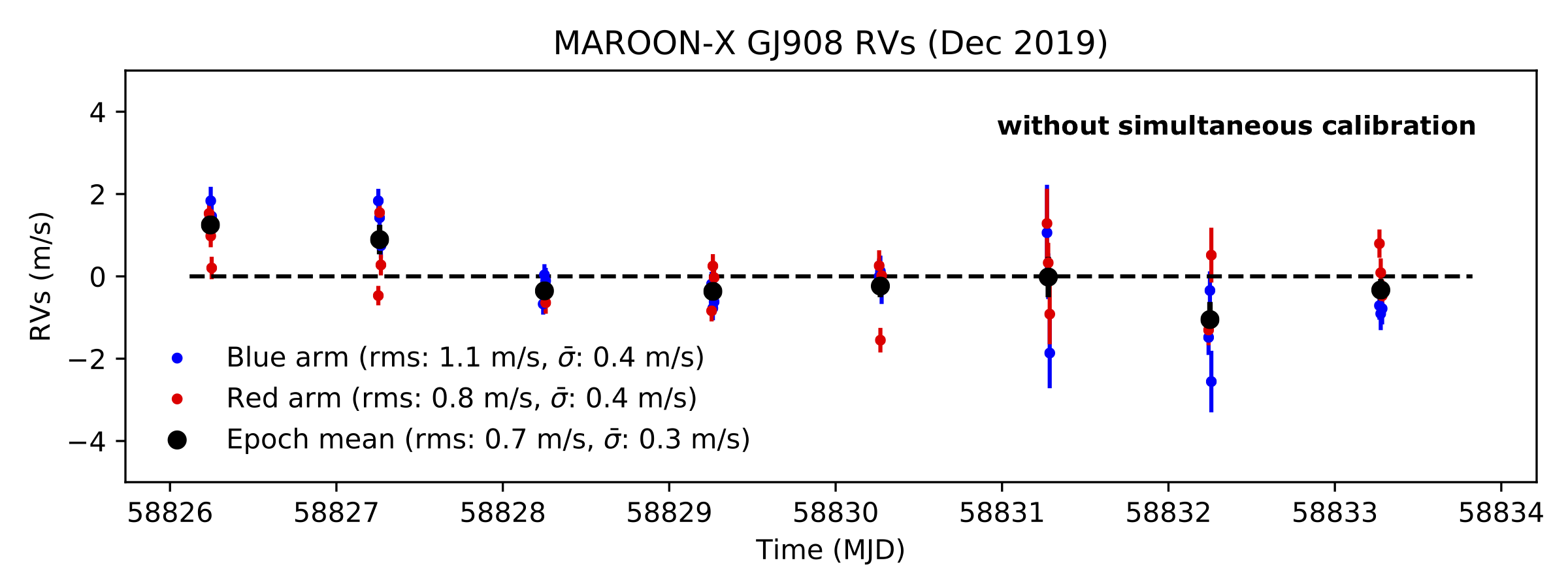}
\includegraphics[width=0.99\textwidth,keepaspectratio]{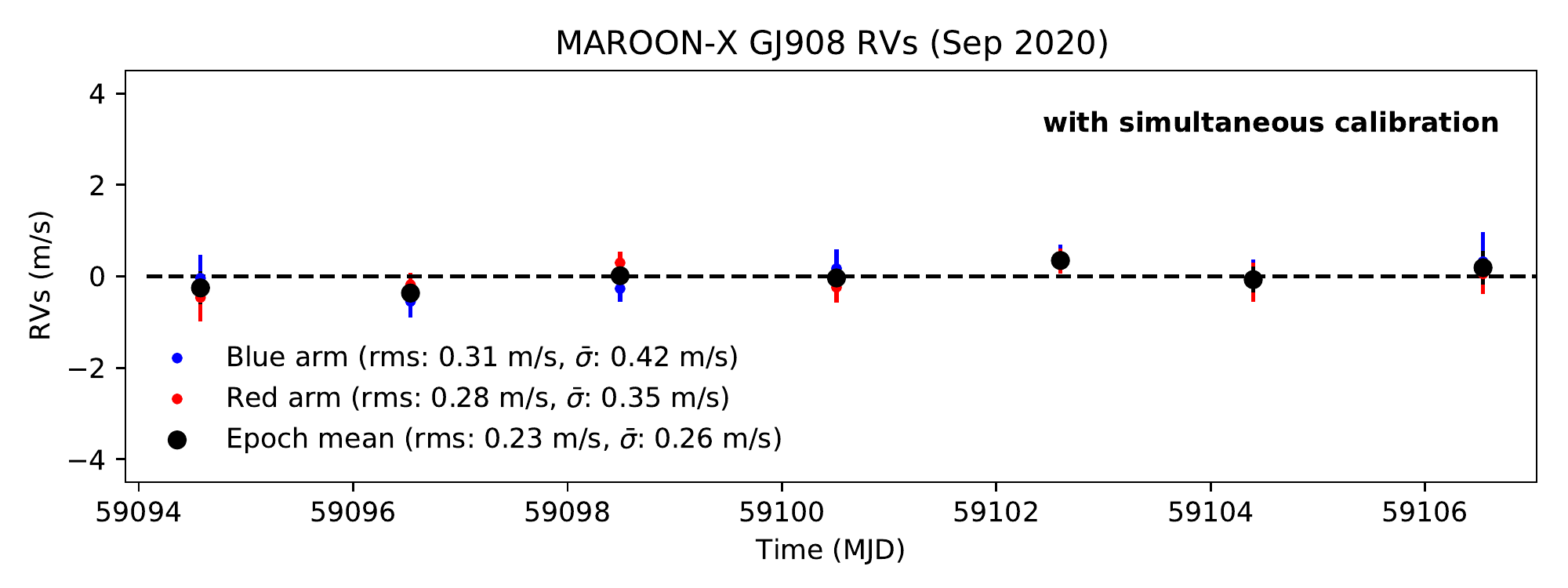}
\caption{\textbf{Radial velocities of the M0.5V RV standard star GJ~908} in December 2019 (top) and in September 2020 (bottom), prior and after the implementation of simultaneous calibrations.}
\label{fig:GJ908}
\end{figure}

During the commissioning and science verification run in December 2019, we started observing a small number of M dwarf stars that we deemed suitable as ``RV standard stars" based on previous observations from other RV surveys. One of the most promising candidates is GJ~908, a M0.5V star selected from the California Planet Survey (CPS\cite{Howard2010}). We took 3--5 back-to-back observations with integration times of 5\,min during 8 consecutive nights. The weather conditions during the run were extremely variable and thus the peak SNR in the red arm varies from 150--650. After analysing the data with \texttt{SERVAL} (see Sec.~\ref{sec:analysis_software}) we obtained radial velocities that show mean internal uncertainties of 0.4\,m\,s$^{-1}$ and a RMS scatter of 0.8\,m\,s$^{-1}$ for the red arm and 1.1
\,m\,s$^{-1}$ for the blue arm. When combining the data from both arms and binning the individual measurements taken typically within 30\,min into a nightly mean, the RMS falls to 0.7\,m\,s$^{-1}$ (see Fig.~\ref{fig:GJ908}). 

The neutral density filter wheel that regulates the intensity of the etalon light coupled into the simultaneous calibration fiber was not operational during the commissioning run. We thus used the ``bracketing technique" and took etalon exposures through the science fibers right before and after each science exposure. It became clear from analysing the etalon time series that instrument drifts were sufficiently non-linear even over short timescales that the specific choices of how to interpolate the etalon data changed the results at the level of up to 50\,cm\,s$^{-1}$ for individual data points and changed the RMS by $\sim$10\,cm\,s$^{-1}$. 

We were able to put the simultaneous calibration fiber into operation right after the commissioning run and all subsequent data are taken with the drift correction directly inferred from the science frames. We find that when using different etalon master calibration frames for the absolute drift measurement, the computed instrument drift is identical to better than 10\,cm\,s$^{-1}$, a significant improvement over the results obtained in the lab in 2017. When we revisited GJ~908 in September 2020, we find the RMS had dropped to $\simeq$30\,cm\,s$^{-1}$ for 7 datapoints taken over a timespan of 14 nights (see lower panel in Fig.~\ref{fig:GJ908}). Combining the blue and red arms improves this value to 23\,cm\,s$^{-1}$. 

\begin{SCfigure}[][t!]
\includegraphics[trim={0cm 0.0cm 0cm 0.0cm},width=0.70\linewidth,clip]{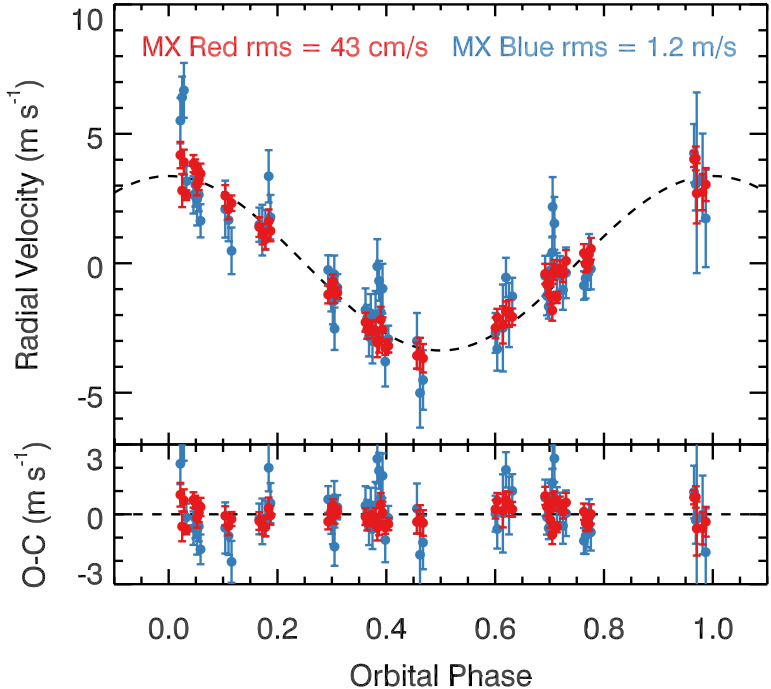}
\caption{{\bf MAROON-X radial velocity measurements of Gl~486b, a transiting rocky planet around a mid-M type star}. This planet was originally identified as a planet candidate from CARMENES radial velocities, and then observed to transit by \textit{TESS}. We performed intensive radial velocity measurements with MAROON-X in semester 2020A to precisely determine the mass and orbit of the planet. No de-trending against or fitting of stellar activity was applied to this dataset.}
\label{fig:wolf437}
\end{SCfigure}

After the temporary shutdown of the Gemini Observatory in the spring of 2020, we started regular science observations in May 2020. Our first scientific result is the confirmation, mass, and orbit of Gl~486b, a transiting rocky planet around a mid-M type star (see Fig.\ref{fig:wolf437}). The target was originally identified as a planet candidate from CARMENES radial velocities and observed to transit with \textit{TESS} in the spring of 2020. We performed intensive radial velocity observations of this planet in May 2020, which lead to a mass measurement better than 5\% (limited by the uncertainty in the mass of the host star) and a 1$\sigma$ error on the predicted secondary eclipse time of only 12 minutes\cite{Trifonov}. Data from the red arm of MAROON-X have a RMS of 44\,cm\,s$^{-1}$, significantly outperforming other RV instrument observing this target. When combining the blue and red arms and binning the data over 30 minutes the RMS falls to $<$30\,cm\,s$^{-1}$.

The results on this science target and our RV standard GJ~908 have been achieved without any de-trending against activity indicators like chromatic index, or differential linewidth, and without fitting and subtracting stellar activity signals as has been done, e.g., for the ESPRESSO observations of Proxima Cen b\cite{Proxima}. The residuals achieved with MAROON-X are thus representing an upper limit on the instrumental stability as they still contain an unknown level of stellar activity.

\section{Future Interventions and Upgrades}
We have planned a number of interventions and hardware upgrades for 2021 to further improve the performance of MAROON-X. These are, in order of decreasing priority: 
\begin{enumerate}
  \item Re-commissioning of the rubidium reference system of the FP etalon calibrator in order to provide long-term stability for RV drift measurement. 
  \item Active temperature control of the Pier Lab and improved temperature control settings for the spectrograph to decrease the absolute drift of MAROON-X.
  \item Re-alignment of the frontend optics to improve the pupil coupling in order to minimize FRD effects and increase the overall throughput of MAROON-X.
  \item Installation of a solar calibrator to feed disk-integrated sunlight to the spectrograph. 
\end{enumerate}

\acknowledgments 
 
The University of Chicago group acknowledges funding for this project from the David and Lucile Packard Foundation through a fellowship to J.L.B., as well as support from the Heising-Simons Foundation, and from The University of Chicago.


\end{document}